\documentclass[aps,prl,twocolumn,10 pt,superscriptaddress,showpacs]{revtex4} 
\usepackage{amssymb}
\usepackage{graphicx}

\usepackage[american]{babel}

\begin{document}

\title{Hysteresis in the electronic transport of V$_2$O$_3$ thin films: non-exponential kinetics and range scale of phase coexistence.}
\author{C.~Grygiel, A.~Pautrat, W.~Prellier and B.~Mercey}
\affiliation{Laboratoire CRISMAT, UMR 6508 du CNRS, ENSICAEN et Universit$\acute{e}$ de Caen, 6 Bd Mar$\acute{e}$chal Juin, F-14050 Caen 4.}


\begin{abstract}
The thermal hysteresis of the electronic transport properties were studied for V$_2$O$_3$ thin films.
The temporal evolution of the resistance shows the out-of-equilibrium nature of this hysteresis with a very slow relaxation. Partial cycles
reveal not only a behavior consistent with phase coexistence, but also the presence of spinodal temperatures which are largely separated. 
The temperature spreading of phase coexistence is consistent with the bulk phase diagram in the pressure-temperature plane, 
confirming that the film is effectively under an effective pressure induced by the substrate.  
\end{abstract}

\maketitle

\section{Introduction}

The physical properties of V$_2$O$_3$, cited as a prototype of strongly correlated systems, have been extensively studied especially for its spectacular
 metal-to-insulator (M-I) transition \cite{foex46,yethiraj90,rice70}. Experimental evidences of a strong pressure dependence of physical properties have
 been also accumulated. Consequently, the phase diagram in the pressure-temperature (P,T) plane was established \cite{whan73}.
A lot of work has been devoted to the M-I Mott transition which can be observed when V$_2$O$_3$ is subjected to a chemical pressure by Chromium substitution \cite{chrome}. Such a negative pressure can be approached in thin films when growing $V_2O_3$
 under tensile stress on LiTaO$_3$ substrate \cite{tensile}.
 The other M-I transition is associated to a change from a rhombohedral symmetry
 to a monoclinic symmetry and occurs at $T\approx$ 160\,{K} under atmospheric pressure \cite{dernier70}. This structural transition leads 
to an increase of the unit-cell volume with range going from $1 \%$ to about $3.5 \%$ \cite{pressurebulk}. This transition has a strong 
first order character,
 and a 10\,{K} thermal hysteresis is often reported. When a hydrostatic (or uniaxial) pressure is applied, the transition temperature 
decreases, and above a critical pressure of roughly 26\,{KBars} the metallic state is completely stabilized down to the
 lowest temperature \cite{pressurebulk}. As a consequence, this strong pressure-dependence of the M-I transition is a motivation for thin film studies since large effective pressure can 
be expected by substrate-induced strains \cite{merwe}. 
Recently, thin films of V$_2$O$_3$ were grown on sapphire substrates, using pulsed laser deposition technique \cite{autier,allimi2008}.
 Under certain growth conditions, the M-I transition of the films can be suppressed at the macroscopic scale for peculiar thicknesses \cite{clarat}.
 Instead of the insulating-like resistance for $T\leq$ 160\,{K}, a broad thermal hysteresis surrounding a non-monotonic (dR/dT) is indeed observed.
 To explain the (dR/dT) shape, we demonstrated that
below 160\,{K}, the samples exhibit an electronic phase separation between metallic and insulating mesoscopic zones
 whose concentration ratio depends on the thickness \cite{clara}.
 Indications of a broad distribution of critical temperatures have also been highlighted and could be a cause for a large hysteresis.
 It is thus necessary to understand this distribution, and consequently its role in the shape of the hysteresis.
In this paper, we show using the technique of minor loops that this broad phase coexistence is responsible for the hysteresis and for non equilibrium transport 
properties over a large temperature scale. 
 
\section{Experimental}

Briefly, a series of  V$_2$O$_3$ thin films, with various thicknesses ($t$), was epitaxially deposited on
  (0001)-sapphire substrate using the pulsed laser deposition technique. Details of the growth can be found elsewhere \cite{clarat}. 
Usually, thin films are under an effective pressure below a critical thickness, 
where the substrate causes the film to grow with uniform elastic strain \cite{matthew}. In the case of a large mismatch between film
 and substrate in-plane lattice parameters, the strain is quickly released in the film, leading to the so-called misfit dislocation regime.
 No critical thickness is then observed down to the ultra-thin film limit.
 It is the case here where the lattice mismatch between V$_2$O$_3$ and sapphire is about 4 \%, as confirmed by X-ray diffraction.
 The tendency is even to recover the bulk parameters at low thicknesses \cite{clara}.
 
The electronic resistivity ($\rho$) is measured using the four probe geometry 
in a commercial PPMS (Physical Properties Measurement System). The resistivity was estimated 
 with the thin rectangular slice approximation:
$\rho = R \pi / ln2 .t R_1(W/L)$, where $R$ is the resistance, $W$ the width, $L$ the length and $R_1(W/L)$ is the correction due to the finite size \cite{tabul}.
A microbridge, whose preparation is described in \cite{clara}, from a 11\,{nm} thick film with a size of $100 \times 300 \mu m^2$ was also measured. The detailed analysis
 of partial cycles and of resistance relaxation were performed in this microbridge. For large current, the voltage-current characteristics
 show non linear behavior, as sometimes observed in samples with phase separation \cite{nonlinear}. However, we limit our study well inside the ohmic regime which is within the range $I=$ 1-500\,{$\mu $A}. We also checked
 that the relaxation of the resistance was uncorrelated with the temperature-dependence of the resistivity and that Joule heating can be avoided.
The results presented hereafter were measured for a typical current of 1\,{$\mu $A}.

\section{First-order transition in V$_2$O$_3$ thin films}

\begin{figure}[!]
\begin{center}
\includegraphics*[width=8.8cm]{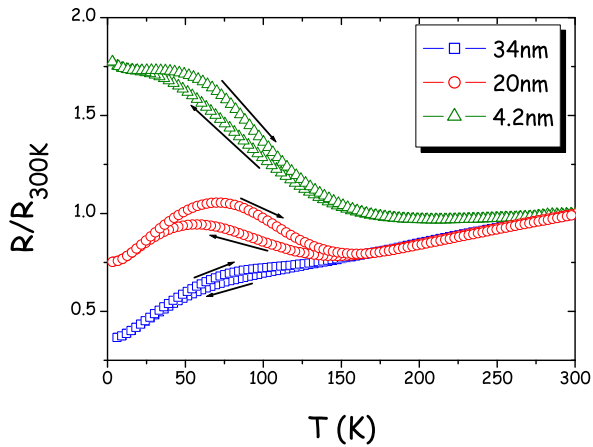}
\end{center}
\caption{Temperature-dependence of the normalized resistance measured during a thermal cycle (cooling-heating, rate=0.5\,{K/min}) for 3 typical thicknesses of 
V$_2$O$_3$ films. A large thermal hysteresis is observed from 20\,{K} $\leq$ T $\leq$ 150\,{K}.}
\label{fig1}
\end{figure}

Recently, we have shown a strong thickness-dependence in the properties of the V$_2$O$_3$ thin films \cite{clarat}.  In particular, we have observed that the M-I transition can be tuned by changing
 the thickness of the films from 2\,{nm} ($\pm $1\,{nm}) up to 300\,{nm} ($\pm $5\,{nm}). 
For example, the M-I transition is observed for ultrathin samples ($t \leq$ 10\,{nm}) with lattice parameters close to the bulk values, whereas thicker samples ($t \geq$ 22\,{nm}) present a quasi-metallic behavior down to low temperatures.
For intermediate thicknesses, the situation is more complex : the films exhibit an intermediate behavior with a clear maximum in the $R(T)$. All samples present an thermal hysteresis $\Delta T$ in the range
 20\,{K}-150\,{K}, but the magnitude of the hysteresis is maximum for the intermediate thicknesses.
These various behaviors are illustrated in figure \ref{fig1} where $R(T)$ curves are plotted for three typical thicknesses.

\begin{figure}[!]
\begin{center}
\includegraphics*[width=8.8cm]{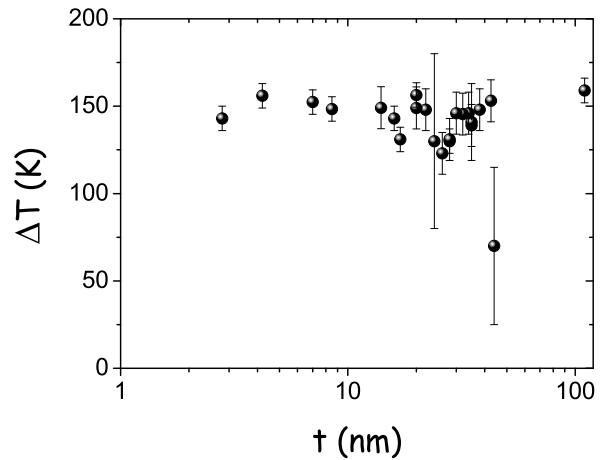}
\end{center}
\caption{Evolution of the thermal hysteresis width as a function of the V$_2$O$_3$ film thickness. The constant trend confirms that no size effect is observed.}
\label{fig2}
\end{figure}

The M-I transition in V$_2$O$_3$ single crystals occurs with strong first order characteristics \cite{firstorder}. Generally, the hysteresis of a first order phase transition results from the possibility of
 supercooling (superheating) metastable states \cite{binder}. The hysteresis observed in our films could indicate such a first order transition, but its width is surprisingly large
 ($\Delta  T \approx  $ 130\,{K} compared to 10\,{K} for the bulk samples).
The broadening of the hysteresis seems to be specific to the thin film form, and may originate from several factors as described below.
The first factor is that the decreasing of the system dimensions (or system thickness) can induce a finite size effect, which can lead to the so-called "rounding field" \cite{round}. 
As a consequence, the width of the first order transition is strongly dependent on the thickness, and $\Delta T_c \propto t^{-1}$. This has
 been indeed used to explain the thermal broadening of the M-I transition in VO$_2$ films \cite{roundVO2}. In our V$_2$O$_3$ samples,
we do not observe any significant thickness-dependence of the thermal hysteresis width as seen in figure \ref{fig2}. 
The second possibility is that the growth generates quenched disorder, which can modify a first order transition
 to a continuous one. Actually, this has been shown for a dimensionality $D=2$ on the basis of general grounds \cite{disorder}. 
For $D=3$, the quenched disorder can also smooth the transition even if this latter still retains first order characteristics. 
To understand the role of such disorder upon the width $\Delta T$ of the hysteresis, our samples were annealed under argon atmosphere, since 
such thermal treatment is usually used to order a sample (as in manganites \cite{prellier}). 
It is unlikely that the quenched disorder (cationic or oxygen disorder) plays a strong role in our films because, even if the residual resistivity
slightly changed by about 10 $\%$, the shape of the $R(T)$ and the width $\Delta T$ of the hysteresis remain unchanged after thermal treatement.
 Finally, a third source of heterogeneity in our films can be the clamping of the unit cell that can induce (likely heterogeneous) strains \cite{strains}.
 Using a two-parallel resistor model, we explained that the peculiar shapes of the $R(T)$ curves (or the non-monotonic thermal variation of the resistance slope $dR/dT$) obtained when cooling the sample arises 
from the coexistence of metallic and insulating zones in the sample \cite{clara}. However, the link between this phase coexistence and the large hysteresis remains unexplored.
The large temperature-range of phase coexistence can indicate a distribution of critical temperatures at a macroscopic scale. Since the reduction of critical temperature is directly correlated with the pressure and
considering that the substrate acts as a strain field, this means that some areas of the film are strongly strained, whereas other ones are partially relaxed. 
In other words, there should be locally a distribution of the pressure within the films. We will now focus on the nature of the thermal resistance hysteresis and show how it can be correlated on the (P,T) phase diagram.

\section{Time-dependent effects and non exponential kinetics}

\begin{figure}[!]
\begin{center}
\includegraphics*[width=8.8cm]{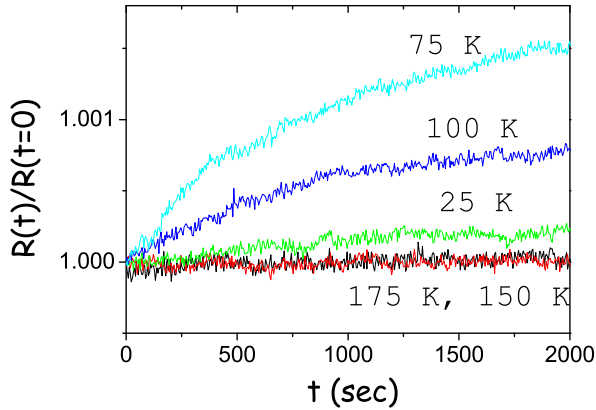}
\end{center}
\caption{Normalized resistance measured as a function of time at different temperatures obtained during a cooling on the microbridge. The time dependence evidences a metastable character.}
\label{fig3}
\end{figure}

To understand the nature of the hysteresis, we have measured on a microbridge of 11\,{nm} thick film the time-dependence of resistance. A slow relaxation can be observed for 20 $K\leq T\leq$ 150 $K$ (Fig. \ref{fig3}).
 The sign of the relaxation depends on whether
 the metastable state is stabilized after heating or cooling the sample.

\begin{figure}[!]
\begin{center}
\includegraphics*[width=8.8cm]{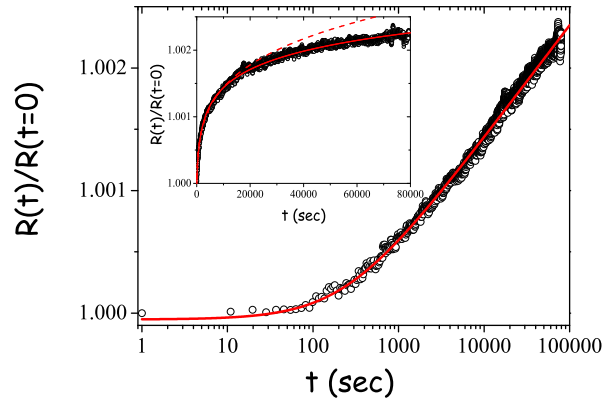}
\end{center}
\caption{Evidence of the non-exponential kinetics of the resistance relaxation at T=80K (The total acquisition time of about one day).
 The data are presented in a semi log-scale to emphasize the logarithm relaxation. The solid line is a fit $R(t)/R(t=0)=1+A.ln(t+\tau _0)$ ($A=0.0004$,$\tau_0=250$). 
In the inset are shown the stretched exponential fit (dashed line) and the logarithm fit of the low time data.}
\label{fig4}
\end{figure}

To fit the time-dependence resistance ($R(t)$), different functional forms can be proposed : exponential, stretched exponential or logarithmic time dependence.
Here, the relaxation is clearly non-exponential. The simplest explanation of such non-exponential kinetics, which is ubiquitous in disordered systems, is that
the relaxation can be intrinsically exponential in different domains, but that the relaxation time is not unique at the sample scale.
 This is in agreement with the idea that different domains with different critical temperatures are present in the sample.
 For non interacting domains, a large statistical averaging of exponential relaxation times leads to the stretched exponential in the form
 $R(t) \propto exp(\pm t/\tau_0 )^\beta $ (where $\tau_0$ is a relaxation time and $\beta < 1$ the relaxation exponent)\cite{stretched}.
 If the acquisition time is moderate (typically of the order of one hour),
 then such a stretched exponential provides a good fitting of the date.
The non-exponential time dependence obtained ($\beta\approx 0.2$) confirms a non Debye relaxation, and indicates that the distribution function of characteristics time is rather flat.
The second case to describe the data leads to a logarithm relaxation of the form $R(t) \propto ln(t+\tau_0)$.
In order to distinguish between the two fits, very long relaxation times (t $\approx$  1 Day) were performed, and the resulting curves are presented in figure \ref{fig4}. The plots show that the relaxation is even more slower than expected from the stretched exponential.
The plot clearly evidences a logarithmic-dependence of the resistivity as a function of time, whereas the stretched exponential is not satisfactory for long periods (unless we do not add a second stretched
 exponential somewhat arbitrarily to fit the data).

Logarithm relaxation is usually observed in complex and strongly interacting materials \cite{stretched}. 
For example, it is observed in glasses, whether they are structural, magnetic or electronic, and this relaxation corresponds to the ageing of the system.
To check the hypothesis of glassy properties in our sample, we tried to observe specific properties of glassy states such as memory and rejuvenation \cite{vincent}.
Note that in the case of a glassy state, a memory effect is usually seen after a negative temperature cycling, 
because the spent time at the low temperature does not influence on the ageing of the system. Here, a negative temperature cycling during the relaxation (80 $K$ $\rightarrow$ 75 $K$ $\rightarrow$ 80 $K$)
 stops the resistance relaxation and then helps the system to reach its equilibrium.
 Since we do not observe genuine glassy characteristics, the slow kinetics are not due to the fact that the system always explores metastable states.
 The relaxation is rather due to the nucleation of M-I domains with a large number of transition temperatures into a M matrix (when cooling the sample).
 This broad distribution of transition temperatures is the reason for the large number of relaxation times inferred by the logarithmic relaxation. This relaxation is very slow, showing that the kinetics are extremely constraint.
 We think that this is due to the long range strain effect induced by the substrate because of the clamping of the unit cell at the bulk transition temperature.
This kind of relaxation is similar to the kinetics of phase transformations
 subjected to pressure \cite{loga}.

\section{Minor loops and phase coexistence}

It is now necessary to determine the region of phase coexistence and the stability limit of the metastable states, e.g. the spinodal limit. 
A way to confirm and to get more informations on the phase coexistence is to perform subloops in the hysteresis cycle. The reason is the following. As far as the
 transformation from one state to another is not complete, partial hysteretic cycles can be observed. 
 
\begin{figure}[!]
\begin{center}
\includegraphics*[width=8.8cm]{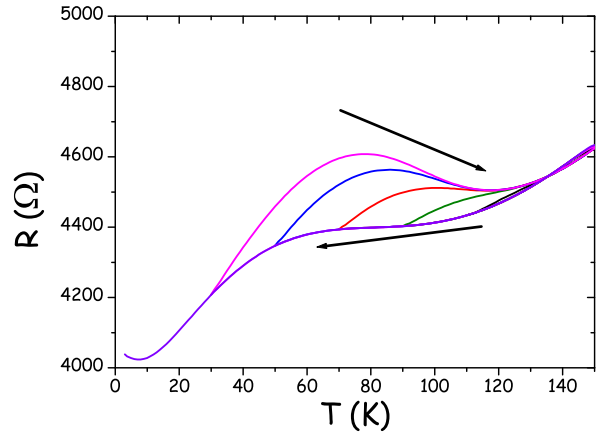}
\end{center}
\caption{Partial cycles of resistance versus temperature with the sequence high T-low T-high T measured on the microbridge. See text for details.}
\label{fig5}
\end{figure}

\begin{figure}[!]
\begin{center}
\includegraphics*[width=9cm]{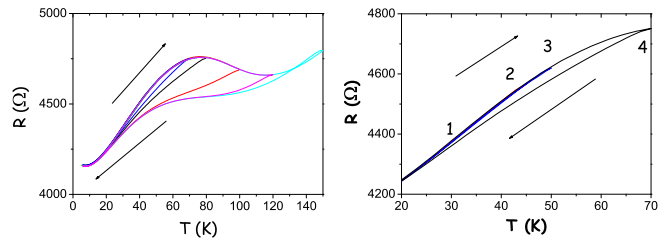}
\end{center}
\caption{Left: Partial cycles of resistance versus temperature with the sequence low T-high T-low T on the microbridge. Right: Zoom of the low temperature partial cycles.
 The return in the cycles is made at the temperature 30$K$ (label 1), 45$K$ (label 2), 50$K$ (label 3), and 70$K$ (label 4) (see text for details).}
\label{fig6}
\end{figure}

\begin{figure}[!]
\begin{center}
\includegraphics*[width=8.8cm]{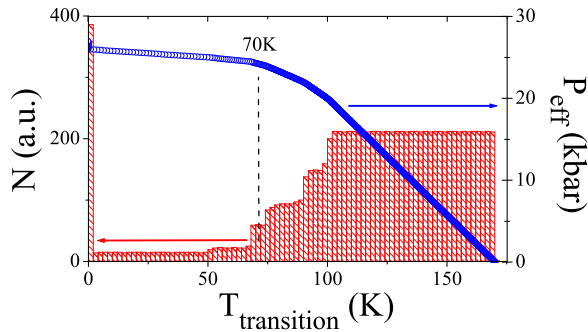}
\end{center}
\caption{Relationship between pressure-temperature (P,T) plane of V$_2$O$_3$ bulk material extracted from \cite{pressurebulk} and a histogram 
of the domain proportion exhibiting one critical temperature (see the text for details).}
\label{fig7}
\end{figure}

Typical partial cycles measured on the microbridge are shown in figure \ref{fig5} and figure \ref{fig6} . 
For the cycling sequence high T$\rightarrow$low T$\rightarrow$high T shown in figure \ref{fig5}, we observe that the metastability begins at $T\approx $ 150\,{K} and 
ends at $T\approx $ 20\,{K}. However, the reverse cycling sequence low T$\rightarrow$high T$\rightarrow$low T shown in figure \ref{fig6} exhibits a quite different behavior.
The three first curves are fully reversible (label 1, 2 and 3 in the figure \ref{fig6}) whereas the 70$K$-one (label 4) shows irreversibility.
Since it is necessary to heat at temperature T $\geq$  70 $K$ to observe that a thermal hysteresis begins to develop, this means that phase coexistence begins at $T\approx $ 70\,{K} and ends at $T\approx $ 150\,{K}. 
This asymmetry appearing for partial cycles and depending on the initial temperature indicates that below 70\,{K}, the high resistance state is strongly favored and can be regarded as the 
stable state (even if some parts of the sample can be stabilized in the low resistance state by supercooling from high temperature). 
Thus, the analysis shows that the zone of phase coexistence extends only from 70\,{K} to 150\,{K}. 
 From the $(P,T)$ bulk phase diagram of the V$_2$O$_3$, the transition line which determines the M-I transition can be described
 as a straight line with a slope $\Delta T_c/ \Delta P \approx -3.5 K/Kbars$ for $T >$ 70\,{K} and a quasi-horizontal line with a slope
 $\Delta T_c/ \Delta P \approx - 40 K/Kbars$ for $T < $ 70\,{K} \cite{pressurebulk} (Fig.\ref{fig7}). 
The pressure-induced at the transition has been thus calculated to be $\Delta P_{max} \approx $ 27\,{KBars} for a complete blocking of the transition \cite{clara}. 
This value is higher than the critical pressure (26\,{KBars}) reported in V$_2$O$_3$ single crystals by McWhan \textit{et al} \cite{pressurebulk} and 
shows that a fully strained film should not exhibit any M-I transition. 
 Since a first order transition leads to an heterogeneous state when the pressure increases at a constant volume, we assume that our film
 sustains a very broad and continuous effective pressure distribution (from $P\approx $ 0 to $P\approx $ 27\,{KBars}, where each pressure is statistically
 equivalent). Therefore, the film should also display a very broad critical temperature distribution ranging from 0\,{K} to 150\,{K}. 
However, from the $(P,T)$ diagram, a simple estimation shows that the phases whose critical temperatures are in the range $0<T_c\leq $ 70\,{K} are statistically unfavorable. 
As a consequence, the temperature range where the phase coexistence occurs is going from 70\,{K} to 150\,{K}, in agreement with the analysis of the partial cycles. This is summarized in figure \ref{fig7}. 
Even if partial cycles were not performed for
 all the thicknesses investigated, the hysteresis cycles were always similar in terms of characteristic temperatures (see Fig.\ref{fig2}).
This suggests that our analysis is not thickness-dependent as far as the width of the phase coexistence is concerned. 
Finally,  one has to note that V$_2$O$_3$ crystals form cracks because of fractures, which occur during the volume change at the M-I transition.
 This is not the case in our films. In bulk V$_2$O$_3$, M and I phases correspond to different structural symmetries,
 which transform into each other via a (local) first order transition. Some indications of the thermoelastic
martensitic effect importance in the M-I transition in V$_2$O$_3$ have been reported using acoustic emission measurements \cite{martensitic1}.
 We propose that in the particular case of V$_2$O$_3$ films, substrate enables a releasing of the strain by inducing a heterogeneous state, as it is the case in a martensitic scenario,
where a large number of small domains increases the spread in transition temperature of
each domain \cite{martensitic2}.

\section{Conclusion}
We have presented a detailed study of thermal hysteresis present in the transport properties of V$_2$O$_3$ thin films. 
Non-exponential relaxation and partial cycles are observed, showing a broad 
 region of phase coexistence. Using partial hysteresis cycles, this zone of the phase coexistence has been precisely established.
The limit of metastability corresponds well to what can be expected if the film is under an effective pressure induced by the substrate.  
We believe that such results will be of the general interest for the thin film physics that display metal-to-insulator transition.

\acknowledgments
 This work was carried out in the frame of the European project CoMePhS (NMP4-CT-2005-517039)
 supported by the European Community and by the CNRS, France.


\begin{references}

\bibitem{foex46}  M.~Foex, C. R. Acad. Sci. \textbf{223}, 1126 (1946).
\bibitem{yethiraj90}  M.~Yethiraj, J. Solid State Chem. \textbf{88}, 53 (1990).
\bibitem {rice70} T.~M.~Rice and D.~B.~McWhan, Metal-Insulator Transition, IBM J. Res. Develop., 251 (1970).
\bibitem{whan73}  D.~B.~McWhan, A.~Menth, J.~P.~Remeika, W.~F.~Brinkman, and T.~M.~Rice, Phys. Rev. B \textbf{7}, 1920 (1973).
\bibitem{chrome} D.~B.~McWhan, and J.~P.~Remeika, Phys. Rev. B \textbf{2}, 3734 (1970).
\bibitem{tensile} S.~Yonezawa, Y.~Muraoka, Y.~Ueda and Z.~Hiroi, Sol. State Comm. \textbf{129}, 245 (2004).
\bibitem{dernier70} P.~Dernier and M.~Marezio, Phys. Rev. B \textbf{2}, 3771 (1970). 
\bibitem{pressurebulk}  D.~B.~McWhan and T.~M.~Rice, Phys. Rev. Lett. \textbf{22}, 887 (1969).
\bibitem{merwe} J.~H.~Van Der Merwe, J. Appl. Phys. \textbf{34}, 117 (1963); J.~H.~Van Der Merwe, J. Appl. Phys. \textbf{34}, 123 (1963).
\bibitem{autier} S.~Autier-Laurent, B.~Mercey, D.~Chippaux, P.~Limelette and Ch.~Simon, Phys. Rev. B \textbf{74}, 195109 (2006).
\bibitem{allimi2008} B.~S.~Allimi, S.~P.~Alpay, C.~K.~Xie, B.~O.~Wells, J.~I.~Budnick, and D.~M.~Pease, Appl. Phys. Lett. \textbf{92}, 202105 (2008).
\bibitem{clarat} C.~Grygiel, Ch.~Simon, B.~Mercey, W.~Prellier, R.~Fr$\acute{e}$sard and P.~Limelette, Appl. Phys. Lett.  \textbf{91}, 262103 (2007).
\bibitem{matthew} J. W. Matthews and A. E. Blakeslee, J. Cryst. Growth \textbf{27}, 118 (1974).
\bibitem{clara} C.~Grygiel, A.~Pautrat, W.~C.~Sheets, W.~Prellier, B.~Mercey and L.~M$\acute{e}$chin, submitted in J. Appl. Phys. and http://arxiv.org/abs/0807.4370.
\bibitem{tabul} The tabulated values can be found in Haldor Topsoe book, "Geometric Factors in Four Point Resistivity Measurement" (1966 revised 1968).
or in F.~M.~Smits, "Measurements of Sheet Resistivity with the Four-Point Probe" , BSTJ, 37, p. 371 (1958). In our case,  $W/L=5$ and $R_1=0.77$.
\bibitem{nonlinear} N. A. Babushkina, L. M. Belova, D. I. Khomskii, K. I. Kugel, O. Y. Gorbenko, and A. R. Kaul, Phys. Rev. B \textbf{59}, 6994 (1999).
\bibitem{firstorder} J.~Feinleib and W.~Paul, Phys. Rev. \textbf{155}, 841 (1967).
\bibitem{binder} K.~Binder, Rep. Prog. Phys. \textbf{50}, 783 (1987).
\bibitem{round} M.~E.~Fisher and A.~N.~Berker, Phys. Rev. B \textbf{26}, 2507 (1982); M.~S.~S.~Challa, D.~P.~Landau and K.~Binder, Phys. Rev. B \textbf{34}, 1841 (1986).
\bibitem{roundVO2} H.~K.~Kim, H.~You, R.~P.~Chiarello, H.~L.~Chang, and T.~J.~Zhang, Phys. Rev. B \textbf{47}, 12900 (1993).
\bibitem{disorder} M.~Aizenman and J.~Wehr, Phys. Rev. Lett.  \textbf{62}, 2503 (1989).
\bibitem{prellier}W.~Prellier, P.~Lecoeur and B.~Mercey, J. Phys. Condens. Matter \textbf{13}, R915 (2001).
\bibitem{strains} Q.~Gan, R.~A.~Rao, C.~B.~Eom, J.~L.~Garret, and M.~Lee, Appl. Phys. Lett. \textbf{72}, 978 (1998) 
\bibitem{stretched} R.~G.~Palmer, D.~L.~Stein, E.~Abrahams, and P.~W.~Anderson, Phys. Rev. Lett. \textbf{53}, 958 (1984).
\bibitem{vincent} E.~Vincent, in Ageing, Rejuvenation and Memory: the Example of Spin-Glasses, Proceedings of the Summer School on Ageing and the Glass Transition, edited by M. Pleimling and R. Sanctuary, Vol. 716 of Lecture Notes in Physics (Springer-Verlag, Berlin, 2007), pp. 7–60.
\bibitem{loga} O.~B.~Tsiok, V.~V.~Brazhkin, A.~G.~Lyapin, and L.~G.~Khvostantsev, Phys. Rev. Lett. \textbf{80}, 999(1998).
\bibitem{martensitic1} F.~A.~Chudnovskii, V.~N.~Andreev, V.~S.~Kuksenko, V.~A.~Piculin, D.~I.~Frolov, P.~A.~Metcalf and J.~M.~Honig, J. Solid State Chem. \textbf{133}, 430 (1997).
\bibitem{martensitic2} V.~N.~Andreev, V.~A.~Pikulin and D.~I.~Frolov, Phys. Solid State  \textbf{42}, 330 (2000).

\end{references}
\end{document}